
\input phyzzx
\hoffset=0.2truein
\voffset=0.1truein
\hsize=6truein
\def\TITLEPAGE{\frontpagetrue}
\def\CALT#1{\hbox to\hsize{\tenpoint \baselineskip=12pt
        \hfil\vtop{
        \hbox{\strut CALT-68-#1}}}}

\def\CALTECH{
        \address{California Institute of Technology,
Pasadena, CA 91125}}
\def\TITLE#1{\vskip .5in \centerline{\fourteenpoint #1}}
\def\AUTHOR#1{\vskip .2in \centerline{#1}}

\def\ABSTRACT#1{\vskip .2in \vfil \centerline{\twelvepoint
\bf Abstract}
        #1 \vfil}
\def\ENDTITLEPAGE{\vfil\eject\pageno=1}

\tolerance=10000
\hfuzz=5pt
\TITLEPAGE
{\hbox to\hsize{\tenpoint \baselineskip=12pt
        \hfil\vtop{\hbox{\strut CALT-68-1940}
        \hbox{\strut CMU-HEP94-22}
        \hbox{\strut DOE RESEARCH AND}
        \hbox{\strut DEVELOPMENT REPORT}
        }}

\TITLE {Strong $\Lambda \pi$ Phase Shifts for CP Violation in Weak
$\Xi \rightarrow \Lambda \pi$ Decay\foot{Work suppported in part by the DOE
under Contract no. DEAC-03-81ER40050 and DOE/ER-40682-76}}
\AUTHOR {Ming Lu and Mark B. Wise}
\CALTECH
\bigskip
\bigskip
\centerline{Martin J. Savage}
\smallskip
\smallskip
\smallskip
\centerline {{\it Department of Physics, Carnegie Mellon University,}}
\centerline {{\it Pittsburgh, PA, 15213}}
\ABSTRACT{Strong interaction $\Lambda\pi$ phase shifts relevant for
the weak nonleptonic decay $\Xi \rightarrow \Lambda \pi$
are calculated using baryon chiral perturbation theory. We find in
leading order that the S-wave phase shift vanishes and the $J={1 \over 2}$
P-wave phase shift is $-1.7 ^{\rm o} $. The small phase shifts imply that CP
violation in this decay will be difficult to observe. Our results follow from
chiral $SU(2)_L\times SU(2)_R$ symmetry. }
\ENDTITLEPAGE
\eject

In the standard six-quark model CP violation arises from the phase $\delta$ in
 the Cabibbo-Kobayashi-Maskawa matrix. So far CP violation has only been
observed in
second order weak $K^0-\bar K^0$ mixing and it is not known if this arises
from the phase $\delta$ or from some new interaction associated
with a very large mass scale. The latter possibility leads to a superweak
scenerio for CP violation [1]. Observation of CP violation in a first order
weak decay amplitude (sometimes referred to as direct CP violation) would rule
out the superweak model as its sole origin. Avenues for detecting CP violation
in first order
weak decay
amplitudes are the measurement of a nonzero value for the
parameter $\epsilon^{\prime}$ and the measurement of asymmetries in $B$ decay.
 Recently it has been proposed to measure  direct CP violation
in the nonleptonic hyperon decay chain $\Xi \rightarrow \Lambda \pi
\rightarrow p \pi \pi$ at Fermilab [2]. To observe CP violation in this case
requires not only a CP violating phase in the weak hyperon decay amplitudes
but also a phase from final state interactions. In this letter we calculate
the strong $\Lambda \pi$ phase shifts that are important for observing CP
violation in $\Xi$ nonleptonic weak decay using baryon chiral
perturbation theory. An interesting aspect of our work is that we are able to
make predictions using only chiral $SU(2)_L \times SU(2)_R$ symmetry
by utilizing the measured $\Sigma^- \rightarrow \Lambda e \bar \nu_e$ decay
rate to determine the magnitude of the $\Sigma \Lambda \pi$ coupling.
Strong phase shifts relevant for weak $\Xi \rightarrow \Lambda \pi$ decay
were first calculated about 30 years ago and we compare our results with these
earlier calculations [3-5]. One important difference between our approach and
the previous work is that we don't rely on $SU(3)$ symmetry. We find that
the S-wave phase
shift vanishes at leading order in chiral perturbation theory and that the
P-wave $J={1 \over 2}$ phase shift is only -$1.7^{\rm o} $. This suggests
that CP violation in the recently proposed hyperon decay experiment at
Fermilab will be  dominated by the $\Lambda \rightarrow p \pi$ part of the
decay chain.

Nonleptonic $\Xi \rightarrow \Lambda \pi$ decay is characterized by S-wave and
P-wave amplitudes which we denote respectively by $S$ and $P$. The
differential decay rate (in the $\Xi$ rest frame) has the form
$$ \eqalign{  {d\Gamma \over d\Omega} \ \ \propto \ \
& 1\
-\  \alpha\ \left(  \hat s_{\Xi} \cdot \hat p_{\pi}\
+\ \hat s_{\Lambda} \cdot \hat p_{\pi}  \right)
\ +\ \beta \ \hat p_{\pi} \cdot \left(\hat s_{\Xi} \times \hat
s_{\Lambda}\right) \cr
& \ \ \  + \ \gamma\  \hat s_{\Xi}\cdot \hat s_{\Lambda}\
+\ (1-\gamma) \left(\hat s_{\Lambda}\cdot \hat p{\pi}\right)\left(\hat s_{\Xi}
\cdot \hat p_{\pi}\right)\ \ \ \ ,} \eqno (1)$$
where $\hat s_{\Xi}$ and $\hat s_{\Lambda}$ are unit vectors along the
direction of the $\Xi$ and $\Lambda$ spins and $\hat p_{\pi}$ is a unit vector
along the direction of the pion momentum. The parameters $\alpha$, $\beta$ and
$\gamma$ which characterize the decay distribution are expressed in terms of
the S-wave and P-wave amplitudes as follows;
$$ \alpha ~=~{2\ {\rm Re}\left( S^*P \right) \over |S|^2+|P|^2} ~, \eqno(2)$$
$$ \beta~=~{2\ {\rm Im} \left( S^*P \right)  \over |S|^2+|P|^2} ~, \eqno(3)$$
$$ \gamma~=~{|S|^2-|P|^2 \over |S|^2+|P|^2} ~. \eqno(4)$$
The S-wave and P-wave amplitudes are complex numbers
$$ S~=~|S|e^{i\delta_S}~~~~,~~~~P~=~|P|e^{i\delta_P}~. \eqno(5)$$
In terms of the modulus and phase of $S$ and $P$ the parameters $\alpha$ and
$\beta$ are
$$\alpha~=~2{|S||P|\over |S|^2+|P|^2}\cos(\delta_S-\delta_P)~, \eqno(6)$$
$$\beta~=~-2{|S||P|\over |S|^2+|P|^2}\sin(\delta_S-\delta_P)~. \eqno(7)$$
Isospin symmetry ensures that $S$ and $P$ for the decays $\Xi^- \rightarrow
\Lambda \pi^-$ and $\Xi^0 \rightarrow \Lambda \pi^0$ are related by a factor
of $\sqrt{2}$.
The quantities $\delta_S$ and $\delta_P$ are respectively equal
(up to a factor of $\pi$) to the strong interaction S-wave and $J={1 \over 2}$
P-wave $\Lambda \pi$ phase shifts plus small but important
contributions from direct weak interaction CP violation [6].

The decay distribution for $\bar \Xi \rightarrow \bar \Lambda \pi$ is also
given by eq.(1) with parameters $\bar \alpha$, $\bar \beta$, $\bar \gamma$
whose expression in terms of the S-wave and P-wave amplitudes $\bar S$ and
$\bar P$ are similar to eqs.(2), (3) and (4). The only difference is that the
analog of eqs.(2) and (3) have a minus sign.
The $\Lambda$'s produced in the decay of unpolarised $\Xi$'s have a
polarisation $\alpha$ (as seen in eq.(1)) and an important measure of CP
violation is the asymmetry
$${\cal A}~=~{\alpha+\bar \alpha \over \alpha -\bar \alpha} ~. \eqno(8)$$
In terms of the phases of the S-wave and P-wave amplitudes the above becomes
$${\cal A}~=~{\cos (\delta_S-\delta_P)-\cos (\delta_{\bar S}-\delta_{\bar
P})\over \cos (\delta_S-\delta_P)+\cos (\delta_{\bar S}-\delta_{\bar P})}~.
\eqno(9)$$

We denote the $J={1 \over 2}$ S-wave and P-wave $\Lambda \pi$ phase shifts by
$\delta_0$ and $\delta_1$ respectively and write
$$ \delta_S-\delta_P~=~\delta_0-\delta_1+\phi_{CP}+\pi~, \eqno(10a)$$
$$ \delta_{\bar S}-\delta_{\bar P}~=~\delta_0-\delta_1-\phi_{CP}+\pi~,
\eqno(10b)$$
with $\phi_{CP}$ the phase that results from direct weak interaction CP
violation. Data from $\Xi^- \rightarrow \Lambda \pi^-$ decay and
$ \Xi^0 \rightarrow \Lambda \pi^0$ decay give (neglecting CP violation)
$\delta_0-\delta_1=(8 \pm 8)^{\rm o}$ and $\delta_0-\delta_1=(38 ^{+12}_{-19}
)^{\rm o}$ respectively. Putting eqs.(10) into eq.(9) yields
$${\cal A}\ = -\tan \phi_{CP} \cdot \tan (\delta_0-\delta_1)~. \eqno(11)$$
Eq.(11) indicates that the CP violating observable ${\cal A}$ is small if the
difference of phase shifts $\delta_0-\delta_1$ is small. In this letter we
calculate the strong interaction $\Lambda \pi$ phase shifts $\delta_0$ and
$\delta_1$ using chiral perturbation theory.

In chiral perturbation theory the pions are encorporated into the $2\times 2$
special unitary matrix
$$ \Sigma~=~exp(2iM/f)~, \eqno(12)$$
where
$$M~=~ \pmatrix{\pi^0/\sqrt 2&\pi^+\cr
\pi^-&-\pi^0/\sqrt 2\cr}~,\eqno(13)$$
and $f\simeq 132~$MeV is the pion decay constant. Under chiral $SU(2)_L\times
SU(2)_R$ symmetry
$$\Sigma \rightarrow L\Sigma R^{\dagger}~,\eqno(14)$$
where $L\in SU(2)_L$ and $R \in SU(2)_R$. It is also convenient to introduce
the square root of $\Sigma$,
$$\xi~=~exp(iM/f)~, \eqno(15)$$
which transforms under $SU(2)_L\times SU(2)_R$ as
$$\xi \rightarrow L \xi U^{\dagger}~=~U \xi R^{\dagger}~.\eqno(16)$$
Here $U$ is a complicated nonlinear function of $L$ , $R$ and the pion fields
$M$. The combinations of meson fields
$$(A^{\mu})^b_a~=~{i \over 2}(\xi \partial^{\mu} \xi^{\dagger}- \xi^{\dagger}
\partial^{\mu} \xi)^b_a ~,\eqno(17a)$$
$$(V^{\mu})^b_a~=~~{i \over 2}(\xi \partial^{\mu} \xi^{\dagger}+ \xi^{\dagger}
\partial^{\mu} \xi)^b_a ~,\eqno(17b)$$
play an important role in the interactions of pions with other fields. Note
that $V^\mu$ contains terms with an even number of pion fields and $A^\mu$
contains terms with an odd number
of pion fields.

The baryon fields we use are the spin ${1\over 2}$ isosinglet $\Lambda$, the
spin ${1\over 2}$ isotriplet $\Sigma_{ab}$
(~$\Sigma_{11} = \Sigma^+\ ,\ \Sigma_{12}\ =\ \Sigma_{21}\ = \
{1\over\sqrt{2}}\Sigma^0$ and $\Sigma_{22} \  = \ \Sigma^-$)
and the spin ${3\over 2}$ isotriplet $\Sigma^{*\mu}_{ab}$ (the assignment of
the $\Sigma^*$'s to $\Sigma^{*\mu}_{ab}$ is analogous to the assignment of the
$\Sigma$'s to $\Sigma_{ab}$).
Under chiral $SU(2)_L\times SU(2)_R$ these fields transform as
$$\Lambda \rightarrow \Lambda~, \eqno(18a)$$
$$\Sigma_{ab}^{(*\mu)} \rightarrow U_{ac} U_{bd}
\Sigma_{cd}^{(*\mu)}~,\eqno(18b)$$
where repeated roman indices $a$,$b$,... are summed over 1,2.
Strong interactions of these baryons with pions are described by a chiral
Lagrangian that is invariant under parity and chiral $SU(2)_L\times SU(2)_R$
symmetry. Expanding in derivatives this chiral Lagrangian density is [7,8],
$${\cal L}~=~{\cal L}_{\Lambda}+{\cal L}_{\Sigma}+{\cal L}_{\Sigma^*}+{\cal
L}_{int}~,\eqno(19)$$
where
$${\cal L}_{\Lambda}~=~\bar \Lambda ~iv \cdot \partial ~\Lambda ,\eqno(20a)$$
$${\cal L}_{\Sigma}~=~\bar \Sigma ^{ab} ~iv \cdot \partial ~\Sigma _{ab}+2 \bar
\Sigma ^{ab} ~v \cdot V^c_a ~\Sigma_{cb}+ (m_{\Lambda}-m_{\Sigma}) \bar
\Sigma^{ab} ~\Sigma_{ab} ~,\eqno(20b)$$
and
$${\cal L}_{int}~=~g_{\Sigma \Lambda} ~\bar \Lambda ~S \cdot A^b_a ~\Sigma
_{cb} ~\epsilon^{ac} + g_{\Sigma^* \Lambda} ~\bar \Lambda ~A^b_a \cdot
{}~\Sigma^*_{bc} ~\epsilon^{ac} + h.c. ~.\eqno(20c)$$
The expression for ${\cal L}_{\Sigma^*}$ is similar to eq.(20b). In eqs.(20)
$S$ is the spin operator four-vector, $\epsilon ^{ac}$ is the antisymmetric
tensor, $\epsilon^{11}=\epsilon^{22}=0$, $ \epsilon ^{12}=-\epsilon ^{21}=1$,
and $v$ is the baryon four-velocity.
There are also interaction terms with one derivative involving two
$\Sigma^{(*)}$ fields and an odd number of pions. However these interactions
are not needed for our computation. We treat $m_{\Sigma}-m_{\Lambda}$ and
$m_{\Sigma^*}-m_{\Lambda}$ as small quantities. For power counting purposes
these mass differences are considered to be the same order as a single
derivative.

The magnitude of the couplings $g_{\Sigma^* \Lambda}$ and $g_{\Sigma \Lambda }$
can be determined from experiment. Comparison of the measured
$\Sigma^{*+} \rightarrow \Lambda \pi^+ $ decay width with
$$\Gamma (\Sigma^{*+} \rightarrow \Lambda \pi^+)~=~g_{\Sigma ^* \Lambda}^2\
{1 \over 6\pi }{|\vec p_{\pi}|^3 \over f^2} {m_\Lambda\over m_{\Sigma^*}}
\ \ \ \ \ ,\eqno(21)$$
gives $g_{\Sigma ^* \Lambda}^2~\simeq~ 1.49$. There is a Goldberger-Treiman
type relation that relates matrix elements of the axial current to the
$\Sigma \Lambda \pi$ coupling $g_{\Sigma \Lambda}$. Using the Noether
procedure we find that in chiral perturbation theory matrix elements of the
left-handed current are given by
$$\bar u \gamma^{\mu}(1-\gamma _5)d~=
g_{\Sigma \Lambda}~\bar \Lambda ~S^{\mu}~\Sigma^-~+...~~~.\eqno(22)$$
In eq.(22) the ellipses denote pieces involving other baryon fields, the pion
fields and terms with derivatives. The resulting
$\Sigma ^- \rightarrow \Lambda e \bar \nu_e$ decay rate is
$$\Gamma (\Sigma ^- \rightarrow \Lambda e \bar \nu _e)~=~{G_F^2\over 80\pi ^3}
|V_{ud}|^2 g_{\Sigma \Lambda}^2~(m_{\Sigma }-m_{\Lambda })^5~.\eqno(23)$$
Comparing with the measured $\Sigma ^- \rightarrow \Lambda e \bar \nu_e$ decay
rate yields $g_{\Sigma \Lambda}^2~\simeq
{}~1.44$.

The Feynman diagrams in Fig.1 determine the S-wave and $J={1 \over 2}$ P-wave
$\Lambda \pi$ phase shifts at the leading order of chiral perturbation theory.
As a function of the pion energy in the center of mass frame,
\foot{In heavy baryon chiral perturbation theory the centre of mass frame and
the baryon rest frame coincide.}
$E_{\pi}$, we find the S-wave phase shift to be
$$\delta_0(E_{\pi})~=~0 ~,\eqno(24)$$
and the $J={1\over 2}$ P-wave phase shift to be
$$\delta_1(E_{\pi})~=~-{(E_{\pi}^2-m_{\pi}^2)^{3/2} \over 12 \pi f^2} \cdot
\Bigl[{1 \over 4}{g_{\Sigma \Lambda}^2\over E_{\pi}+m_{\Sigma}-m_{\Lambda}}
+{3 \over 4}{g_{\Sigma \Lambda}^2 \over E_{\pi}+m_{\Lambda}-m_{\Sigma}}$$
$$~~~~~~~~~~~~~~~~~~~~~~~~~~~~~~~~~~~-{4 \over 3}{g_{\Sigma ^* \Lambda}^2\over
E_{\pi}+m_{\Sigma ^*}-m_{\Lambda}}\Bigr] ~.\eqno(25)$$

The expression inside the bracket for $\delta_1 (E_\pi)$ is singular at the
unphysical
pion energy $E_{\pi}=m_{\Sigma}-m_{\Lambda}$ because of the $\Sigma$ pole.
(When
the energy is near this value other terms we have neglected become important
and tame the singularity.)
Note that there is no singularity at $E_{\pi}=m_{\Sigma ^*}-m_{\Lambda }$ as
there is no $\Sigma^*$ pole in the $J={1\over 2}$ channel.
The S-wave phase shift vanished because $\Lambda$ is an isospin-zero baryon
(hence there is no coupling to two pions in ${\cal L}_{\Lambda}$) and because
the $\Sigma \Lambda \pi$ and $\Sigma^* \Lambda \pi$ interactions have the
pions derivatively coupled.  At higher order in chiral perturbation theory
we expect an S-wave phase shift suppressed by a factor of order
$E_{\pi}/\Lambda_\chi$
(where $\Lambda_\chi$ is the chiral symmetry breaking
scale) compared, for example, with the S-wave pion-nucleon phase shifts.
At $E_\pi\sim 200$MeV the pion-nucleon S-wave phase shifts are several degrees.
A contribution to the S-wave $\Lambda\pi$ phase shift suppressed by
$E_\pi/\Lambda_\chi$ arises from higher derivative terms, eg.
$${\cal L}_{\rm higher}\ = \ {c\over\Lambda_\chi}\ \overline{\Lambda}\Lambda\
A^a_b\cdot A^b_a \ \ \ .\eqno(26)$$
The coefficient $c$ is expected to be of order unity, but as it is an unknown
quantity the S-wave phase shift at this order is not calculable.
Previous calculations [5] did not find a small S-wave $\Lambda \pi$ phase
shift. Fig.2 contains a plot of the $J={1\over 2}$ P-wave phase shift
$\delta_1$ as a function of $E_{\pi}$. For the
hyperon decay $\Xi \rightarrow \Lambda \pi$ we need the phase shift evaluated
at $E_{\pi} \simeq m_{\Xi}-m_{\Lambda} = 206 {\rm MeV}$.
At this energy the $J={1\over 2}$ P-wave phase shift is
$\delta_1  = -1.7^{\rm o}$.
This is within a factor of two of the value for $\delta_1$ obtained in
previous calculations [3,5].

Our predictions for the phase shifts do not make use of chiral $SU(3)_L \times
SU(3)_R$ symmetry. However, chiral perturbation theory is an expansion in
$E_{\pi}$ and our result for $\delta_1(m_{\Xi}-m_{\Lambda})$ relies on
$m_{\Xi}-m_{\Lambda}$ and hence the strange quark mass being small compared
with the chiral symmetry breaking scale.

The smallness of the $J={1\over 2}$ P-wave phase shift $\delta_1 (E_\pi)$
is partly the result of a cancellation
between the Feynman diagrams involving the $\Sigma$ and
$\Sigma^*$. This cancellation becomes exact in the large $N_c$ limit [8,9]
where
$m_{\Sigma}=m_{\Sigma^*}=m_{\Lambda}$ and
$g_{\Sigma^* \Lambda}^2={3 \over 4}g_{\Sigma \Lambda}^2$.

Our calculations indicate that for the weak decay
$\Xi \rightarrow \Lambda \pi$ the difference between the S- and $J={1\over 2}$
P-wave phase shifts
$\delta_0-\delta_1$ and consequently the CP violating asymmetry ${\cal A}$ are
small.
Therefore, it is likely that any CP violation observed in the recently
proposed Fermilab experiment will be dominated by CP violation in the
$\Lambda \rightarrow p \pi$ part of the $\Xi \rightarrow \Lambda \pi
\rightarrow p \pi \pi$ decay chain.

\bigskip

MJS thanks the High Energy Physics group at Caltech for their kind hospitality
during the course of this work.

\bigskip

\centerline {\bf References}

\item{1.} B. Winstein and L. Wolfenstein, Rev. Mod. Phys. {\bf 65} (1993) 4,
and references therein.

\item{2.} G. Gidal et al, Fermilab Proposal P871 (1993) (unpublished).

\item{3.} B.R. Martin, Phys. Rev. {\bf 138} (1965) B1136.

\item{4.} E. Abers and C. Zumach, Phys. Rev. {\bf 131} (1963) 2305.

\item{5.} R. Nath and A. Kumar, Nuovo Cimento {\bf 36} (1965) 669.

\item{6.} J.F. Donoghue, X.-G. He and S. Pakvasa, Phys. Rev. {\bf D34} (1986)
833; X.-G. He, H. Steger and G. Valencia, Phys. Lett. {\bf B272} (1991) 411.

\item{7.} E. Jenkins and A.V. Manohar, Phys. Lett. {\bf B255} (1991) 558;
E. Jenkins and A.V. Manohar, in Proceedings of the Workshop on Effective
Field Theories of the Standard Model, Dobogoko, Hungary, Aug. 22-26 (1991),
ed. U.G. Meissner, World Scientific.

\item{8.} J.L. Gervais and B. Sakita, Phys. Rev. Lett. {\bf 52} (1984) 87;
Phys. Rev. {\bf D30} (1984) 1795.

\item{9.} R. Dashen, E. Jenkins and A.V. Manohar, Phys. Lett. {\bf B315} (1993)
425; Phys. Lett. {\bf B315} (1993) 438.

\bigskip
\bigskip

\centerline{\bf Figure Captions}
\smallskip

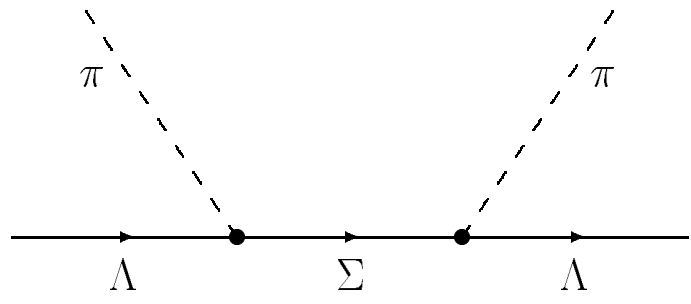
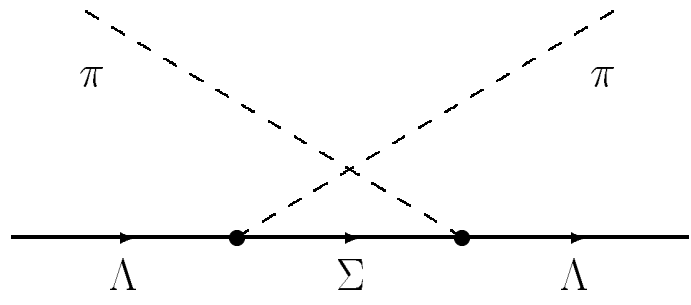
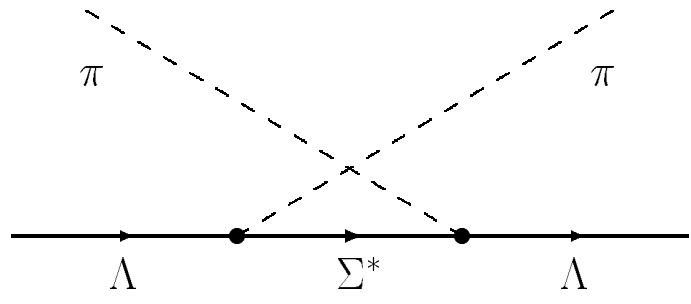
\noindent  Figure 1. \ \ Feynman diagrams contributing to the $J={1\over 2}$
P-wave
$\Lambda\pi$ phase shift $\delta_1$.  There is no contribution to the S-wave
phase shift at leading order in chiral perturbation theory.

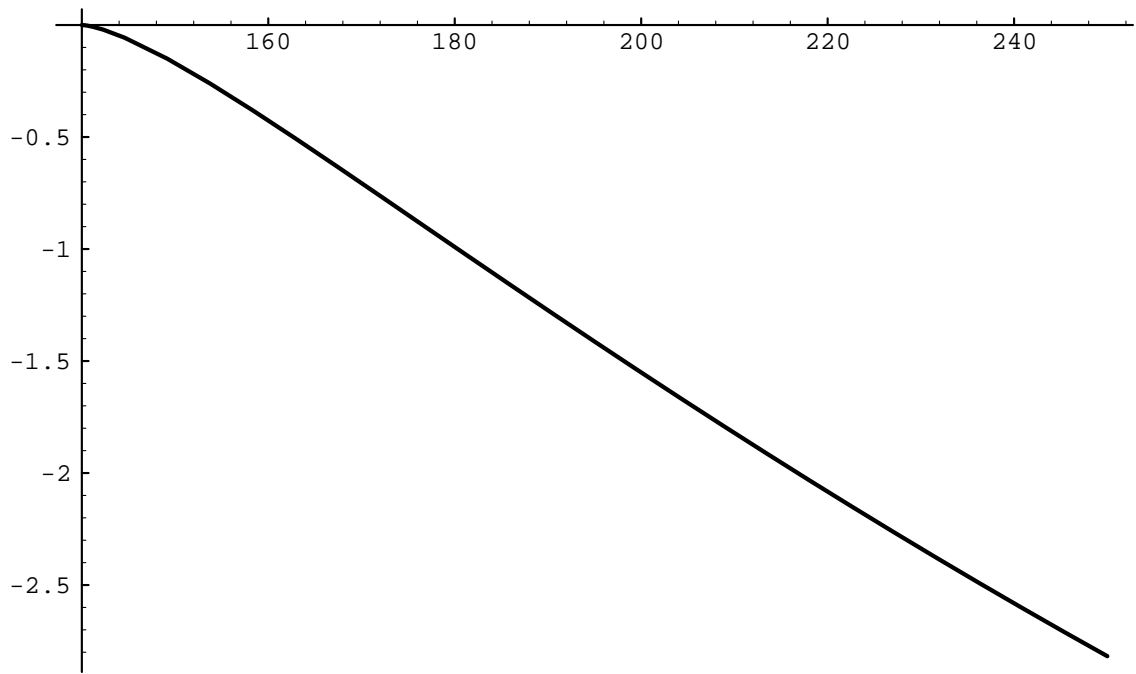
\noindent Figure 2. \ \ The $J={1\over 2}$ P-wave phase shift $\delta_1$ (in
degrees) as a
function of pion energy $E_\pi$ (in MeV).

\end